\documentclass[showpacs,showkeys,11pt,
preprint,preprintnumbers,nofootinbib,
groupedaddress,superscriptaddress,amsmath,amssymb]{revtex4}
\usepackage{amsfonts}
\usepackage{amsmath}
\usepackage{graphics}
\usepackage{epsfig}
\usepackage{url}
\usepackage{multirow}
\usepackage{feynmp}
\usepackage{xcolor}
\newcommand {\be}{\begin{equation}}
\newcommand {\ee}{\end{equation}}
\newcommand {\ba}{\begin{eqnarray}}
\newcommand {\ea}{\end{eqnarray}}

\begin{document}
\title{Charged   Higgs   Observability   via   Charged Higgs   Pair   Production   at   Future   Lepton Collider}

%%%%%%%%%%%%%%%%%%%%%%%%%%%%%%%%%%%%%%%%%%%%%%%%%%%%%%%%%%%%%%%%%%%%%%%%%%%%%%%%

\author{Ijaz Ahmed}
\email{ijaz.ahmed@riphah.edu.pk}
\affiliation{Riphah International University, Sector I-14, Hajj Complex, Islamabad Pakistan}
\author{Nadia Kausar}
\email{nkausar430@gmail.com}
\affiliation{Riphah International University, Sector I-14, Hajj Complex, Islamabad Pakistan}
\author{Wahajaht Sagheer}
\email{Wahjahatsagheer@gmail.com}
\affiliation{Riphah International University, Sector I-14, Hajj Complex, Islamabad Pakistan}
\author{Ather M. W.}
\email{mohsan@nutech.edu.pk}
\affiliation{National University of Technology, Islamabad}
%%%%%%%%%%%%%%%%%%%%%%%%%%%%%%%%%%%%%%%%%%%%%%%%%%%%%%%%%%%%%%%%%%%%%%%%%%%%%%%%
%\date{\today

\begin{abstract}
The observability of charged Higgs $ H^{\pm}  $ has been  investigated at future lepton collider by assuming type-I 2HDM, at a centre of mass energy $ \sqrt{s}=1.5$ TeV. The signal process chain is  $ e^{+}e^{-} \rightarrow Z^{*}/ \gamma^{*}\rightarrow H^{+} H^{-}\rightarrow H W^{+} H W^{-}\rightarrow   b \overline{b} jjb \overline{b} jj$. The process proceed through virtual gamma and Z-boson exchange  in S-channel. Several benchmark points are selected and events are analyzed to reconstruct the  mass of charged Higgs bosons $ H^{\pm} $. The value of $\tan\beta $ is kept relatively high to enhance the branching ratio  of $ H\rightarrow b\overline{b} $ to benefit the signal processes. The main SM background processes  produced is $ t\overline{t} $. Signal selection and  significance efficiencies are  calculated at integrated luminosities of  $100 fb^{-1} $ and $ 500 fb^{-1} $. The  reconstructed and corrected mass of charged Higgs bosons $ H^{\pm} $ is determined. Through analyzing the results, it is demonstrated  that charged Higgs bosons can be  discovered through pair production process  via its bosonic decays. This study is supposed to provide the experimentalists a good way to examine the  Higgs bosons  beyond SM, as well as to check the validity of 2HDM models in considered parameter space.
\end{abstract}

\maketitle
\flushbottom

\section{Introduction}
\begin{figure}[tbp]
\centering 
\includegraphics[width=0.78\textwidth,clip]{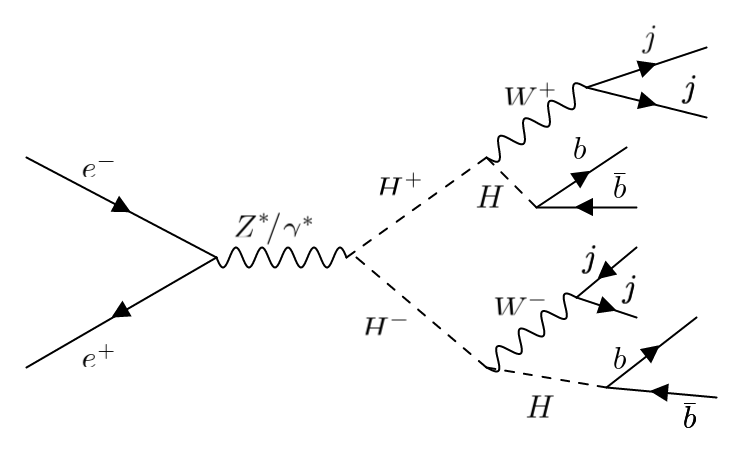}
\caption{\label{fig:MZ} The s-channel Feynman diagram  contributing to the signal process}
\end{figure}

Particle physics is concerned with the study of elementary particles  and their interactions. Many theories were developed to explain the properties and interactions of all known particles. The Standard Model (SM) is founded to be consistent with experimental results till the time.\cite{constraints}. In Standard Model,  only one Higgs doublet exists and the masses of bosons and fermions  are obtained by the Higgs mechanism~\cite{search}. SM have some unanswered questions like the neutrino oscillations, dark matter and the absence of gravity in the theory. These problems made the basis for the development of the theories beyond the Standard Model which may be capable to resolve these mysteries. The most simplest enhancement of Standard Model is the Two Higgs Doublet Model (2HDM). This model has two Higgs doublets and allow more physical Higgs states. Two Higgs Doublet Model (2HDM) is  the simplest low energy effective model~\cite{low} and provides general description of  Higgs sector and its interaction with fermions \cite{constraint}.
In 2HDM the complex scalar doublet  $\phi _{1}$ of SM is augmented with another doublet , $ \phi _{2} $. Due to second doublet all gauge  bosons and fermions  gain mass~\cite{theory}. There are eight degrees of freedom in 2HDM. The three degrees of freedom are eaten up by the electroweak bosons in electro-weak symmetry breaking. The remaining five degrees of freedom give rise to five physical Higss bosons ~\cite{observe}. The 2HDM contains $2$ even  neutral scalar Higgs bosons  $ h$ , $ H $ , an odd pseudoscalar neutral Higgs boson $ A $ and a pair of charged Higgs bosons $ H^{\pm} $ ~\cite{lc}. The presence of charged higgs bosons is a special characteristics of 2HDM because these are not included in SM.\\
With the discovery of Higgs boson at Large Hadron Collider (LHC), in 2012, it was confirmed that the thought of Higgs mechanism is a right approach. This discovery begins a race among the scientists to verify that either there exist only a single observed Higgs or more Higgs states are also possible as predicted by various extended theories. Till date no such boson has been found. Compact Linear Collider (CLIC) is one of the proposed future lepton collider to accelerate the electron and positron beams. CLIC is aimed to be constructed and operated at  three steps, i.e. at $380$ GeV, $1.5$ TeV and $3$ TeV collision energies respectively. Lepton collider can measure the Higgs couplings with more precision than already achieved by LHC.  It is thought that due to its clean environment, CLIC is more capable for finding new physics that is why in this study, the  phenomenology for this collider is made~\cite{CLIC}. The aim of designing CLIC is to achieve the electron-positron head-on collisions at several tera electron volts (TeV) energies.\\The focus of the paper is on the study of charged Higgs pair bosons $H ^{\pm}$ and their bosonic decays. The theoretical framework for this work is type 1 2HDM. The investigation of  the  observability of charged Higgs bosons $H^{\pm}$ takes place by the signal process   $ e^{+}e^{-} \rightarrow Z^{*}/\gamma^{*}\rightarrow H^{+} H^{-}\rightarrow H W^{+} H W^{-}\rightarrow   b \overline{b} jj  b\overline{b} jj$. Where $j$ represents    light jet and b is the bottom quark. This is the most favorable production scenario with bosonic decays for charged Higgs study at linear colliders. The  process will proceed by exchange of   gamma and Z-boson in s-channel. The Figure~\ref{fig:MZ} shows the Feynman  diagram of this process.
In this analysis, each of the charged Higgs decays into $W ^{\pm} H$ 
and the final state becomes  $2W ^{\pm} H$.  Several benchmark points in the parameter space of 2HDM  \cite{majid}   are considered for charged Higgs bosons observability  at centre of mass energy $\sqrt{s}= 1.5$ TeV. The signal and background events are generated independently for each scenario \cite{observe}. The  charged Higgs boson candidates are identified by comparing signal with background events in different distributions. 
To reconstruct the charged Higgs bosons, simulated events are analyzed. For this purpose, first b-jets are identified. These  b-jets are reconstructed by choosing proper  b tagging algorithms and b-jet clustering. The combinations of b-jets are identified so that the dijet invariant mass gives us the Higgs candidate. The mass range for studied charged Higgs is $125 <m_{H^{\pm}}  < 400$  GeV at $\sqrt{s}  = 1.5$ TeV at the integrated luminosity of $500fb^{-1}$
\section{Two Higgs Doublet Model} 
There are four types of  2HDM. In type I, the gauge bosons and all fermions attain mass from one Higgs doublet while contribution of the other Higgs doublet is via mixing~\cite{constraint}.  Discrete Symmetry $ \phi_{1} \rightarrow -\phi_{1} $ is involved in type I. One higgs doublet (conventionally $ \phi_{2} $ ) couples with all quarks (up-type quark and down-type quark) and charged leptons in type-I 2HDM and  flavour is conserved naturally. There are two discrete situations to achieve naturally flavor conservation in 2HDM. First, when only one Higgs doublet (conventionally $ \phi_{2} $ ) couple with all quarks (up-type and down-type) and charged leptons, it is named as the type-I 2HDM. In second case, termed as type-II 2HDM, all the up-type quarks couple to one Higgs doublet ($ \phi_{2} $ ) and all down-type quarks along with charged leptons couple with the other Higgs doublet ($ \phi_{1} $ ). It can be observed that a discrete symmetry $( \phi_{1} \rightarrow - \phi_{1} )$ can be implemented on type-I as well as on type-II. It is conventionally assumed that the up-type quarks are always coupled with the second doublet $  \phi_{2}$ in all types. Different types  of 2HDM are shown in Table ~\ref{tab:1}. 
\begin{table}[tbp]
	\centering
\begin{tabular}{|c|c|c|c|c|}
\hline
Types of Model& Description&$u_{R}^{i}$ & $d_{R}^{i}$ &  $e_{R}^{i}$ \\
\hline
Type I & Fermiophobic&$\Phi_{2}$ & $\Phi_{2}$ &  $\Phi_{2}$ \\
\hline
Type II &MSSM like &$\Phi_{2}$ & $\Phi_{1}$ & $\Phi_{1}$ \\
\hline
Type III&Lepton-specific & $\Phi_{2}$ & $\Phi_{2}$ & $\Phi_{1}$ \\
\hline
Type IV&Flipped & $\Phi_{2}$ & $\Phi_{1}$ & $\Phi_{2}$ \\
\hline
\end{tabular}
\caption{\label{tab:1}Different types of 2HDM on the  basis of coupling of Fermions with charged leptons }
\end{table}
The expression for the general scalar potential with two higgs doublets $ \phi_{1} $ and $ \phi_{2} $ can be written as
\begin{equation}
\label{eq:A}
\begin{aligned}
V_{2HDM}=m_{11}^{2}\phi_{1}^{\dagger}\phi_{1}+m_{22}^{2}\phi_{2}^{\dagger}\phi_{2}
-[m_{12}^{2}\phi _{1}^{\dagger}\phi _{2}+h.c]+\frac{1}{2}\lambda_{1}(\phi _{1}^{\dagger}\phi_{1})^{2}+\frac{1}{2}\lambda_{2}(\phi_{2}^{\dagger}\phi_{2})^{2}\\+\lambda_{3}(\phi _{1}^{\dagger}\phi_{1}) (\phi_{2}^{\dagger}\phi _{2})+
\lambda_{4}(\phi_{1}^{\dagger}\phi_{2})(\phi _{2}^{\dagger}\phi_{1})+[\frac{1}{2}\lambda_{5}(\phi_{1}^{\dagger}\phi_{2})^{2}+[\lambda_{6}(\phi_{1}^{\dagger}\phi_{1})+\lambda_{7}(\phi_{2}^{\dagger}\phi_{2})](\phi_{1}^{\dagger}\phi_{2})+h.c]
\end{aligned}
\end{equation}
In Equation \eqref{eq:A}, $\lambda_{i}$, where ($i=1,2.....,7)$ are  coupling parameters having no dimensions and $m_{11}^{2}$,  $m_{22}^{2}$ and $m_{12}^{2}$ are squared mass parameters. Out of these parameters the $ m_{12}^{2}$ and $\lambda_{i}$ (Where $i = 5,6,7$) are preferably complex while the remaining parameters are real. The $ m_{12}^{2} $ term is of great importance because its nonzero values causes the soft breakdown  of the $ Z_{2} $ symmetry $ (\phi_{1}\rightarrow -\phi_{1}) $ or $ (\phi_{2}\rightarrow -\phi_{2}) $. The potential becomes explicitly CP violating in the presence of non zero imaginary parts of the complex parameters therefore in order to treat the potential as CP conserving here it is supposed that all the parameters used are real. For complete specification of the model  the different parameters  $m_{12}^{2}$, $ \tan\beta $, physical Higgs masses$ m_{H^\pm} $,$ m_{A} $,$ m_{h} $,$ m_{H} $, mixing angle $ \alpha $,$ \sin(\beta - \alpha) $,  $\lambda_{6}$ and $\lambda_{7}$ must be calculated in the physical basis~\cite{2HDM,model}.If the value of $ \sin(\beta - \alpha) $ is taken to be one then it is called the exact alignment limit in which the lighter CP-even Higgs $ h $  behaves like SM Higgs boson and if the value of $ \sin(\beta - \alpha) $ is taken to be zero then the heavier CP-even Higgs behaves like SM Higgs boson. This study is performed within alignment limit i.e. $ \sin(\beta - \alpha) = 1$. The mass of lighter Higgs is taken as the mass of SM higgs boson i.e. $m _h = 125GeV$.
In  type-I 2HDM all the fermions couple with a single Higgs doublet  $(\phi_{2} ) $  same as in  SM however the other Higgs doublet does not couple at all. This is due to the implementation of distinct $Z_{2}  $ symmetry. For Yukawa interactions in  type-I 2HDM the lagrangian is given in Equation~\eqref{eq:m3} 
\begin{equation}
\label{eq:m3}
L_{Yukawa}=Y_{e}\overline{L_{L}}\phi_{2}e_{R}+Y_{u}\overline{Q_{L}}   ~\widehat{\phi}_{2}u_{R}+Y_{d}\overline{Q_{L}} \phi_{2}d_{R}+h.c]
\end{equation}
where $e_{R},_u{R},d_{R}$  are the left handed leptons, up-type and down-type quarks singlet respectively with$ Y_{e},Y_{u},Y_{d} $ as their corresponding Yukawa coupling matrices. The terms $ \overline{L_{L}},\overline{Q_{L}} $ are left-handed lepton and quark doublets respectively and $ \widehat{\phi}_{2}=\iota\sigma_{2}\phi_{2}^{\ast} $ (Where $ \sigma_{2} $ is Pauli matrix). If the weak eigen states of $ \phi_{2} $ are expressed in physical terms then above Equation ~\eqref{eq:m3} becomes
\begin{multline}
\label{eq:m21}
L_{Yukawa}=-\lbrace\underset{\varphi = u,d,l} \sum \dfrac{m_{\varphi}} {\upsilon} (\kappa_{\varphi}^{h} \overline{\varphi} \varphi h^{0}+\kappa_{\varphi}^{H} \overline{\varphi} \varphi H^{0}- \iota\kappa_{\varphi}^{A}\overline{\varphi}\gamma_{5} \varphi A\rbrace- \lbrace\dfrac{ V_{ud} }{\sqrt{2}\upsilon}\overline{u}(m_{u}\kappa_{u}^{A}P_{L} + m_{d}\kappa_{d}^{A}P_{R})dH^{+}\\
 - \dfrac{m_{l}}{\sqrt{2}\upsilon}\kappa_{l}^{A}\overline{\nu}_{L}l_{R}H^{+}+h.c\rbrace
 \end{multline}
In above Equation the $ \kappa $ factors represent the Yukawa couplings and their values for 2HDM type-I are given in Table ~\ref{tab:mz} .
\
\begin{table}[tbp]
\centering
\begin{tabular}{|c|c|c|c|c|c|c|c|c|c|}
\hline
$\kappa_{u}^{h}$&$\kappa_{d}^{h}$&$\kappa_{l}^{h}$&$\kappa_{u}^{H}$&$\kappa_{d}^{H}$&$\kappa_{l}^{H}$&$\kappa_{u}^{A}$&$\kappa_{d}^{A}$&$\kappa_{l}^{A}$\\
\hline
$\dfrac{\cos\alpha }{\sin\beta}$&$\dfrac{\cos\alpha }{\sin\beta}$&$\dfrac{\cos\alpha }{\sin\beta}$&$\dfrac{\sin\alpha }{\sin\beta}$&$\dfrac{\sin\alpha }{\sin\beta}$&$\dfrac{\sin\alpha }{\sin\beta}$&$\cot\beta $&$ -\cot\beta $&$ -\cot\beta $\\
\hline
\end{tabular}
\caption{\label{tab:mz}Values of Yukawa couplings for 2HDM type I}
\end{table}
For all types of 2HDM, W and Z boson couplings with neutral Higgs bosons are identical. Light Higgs ‘h’ and heavy Higgs ‘H’ couplings with either ZZ or WW are equal to $ \sin(\beta-\alpha) $ and $\cos(\beta-\alpha)  $ times the corresponding standard model couplings respectively. 
The coupling for the interaction between higgs boson pair and gauge
boson like $ H^{\pm}W^{\pm}H $ can be estimated by using the gauge
coupling structure and the angles $\alpha$ and $\beta$. The coupling relations \cite{bb} are given below in  Equations ~\eqref{eq:ms},\eqref{eq:ms1} and \eqref{eq:ms2}.
\begin{equation}
\label{eq:ms}
g(H^{\pm} H W^{\pm}):\dfrac{g}{2}(p_{H}-p_{H^{\pm}})^{\mu}\sin (\beta-\alpha)
\end{equation}
\begin{equation}
\label{eq:ms1}
g(H^{\pm} h W^{\pm}):\dfrac{g}{2}(p_{h}-p_{H^{\pm}})^{\mu}\cos (\beta-\alpha)
\end{equation}
\begin{equation}
\label{eq:ms2}
(gH^{\pm} H W^{\pm}):\dfrac{g}{2}(p_{A}-p_{H^{\pm}})^{\mu}
\end{equation}
Where $ p_{H} $, $ p_{h} $ and $ p_{A}$ represent the momentum of incoming particles respectively. 
\section{ Strategy  and Signal Extraction }
In this study events are produced with the help of Pythia-8210 ~\cite{pyt} and relative efficiencies are also calculated in it. 2HDMC-1.7.0 ~\cite{con}is used to calculate the branching ratios as well as decay widths of Higgs sector in type-I of Two Higgs doublet Model at desired benchmark points and its output in SLHA format is fed to Pythia. For the analysis and reconstruction of jets produced in the events Pythia is linked with fastjet-3.3.3. In order to record the events data, the interface of HepMC-2.06.06~ \cite{ Hep} is given to Pythia. The output of Pythia is than analyzed and histograms are plotted by using Root- 6.20/04 ~\cite{rt}.
\subsection{Signal Process}
\begin{figure}[tbp]
\centering 
\includegraphics[width=0.78\textwidth,clip]{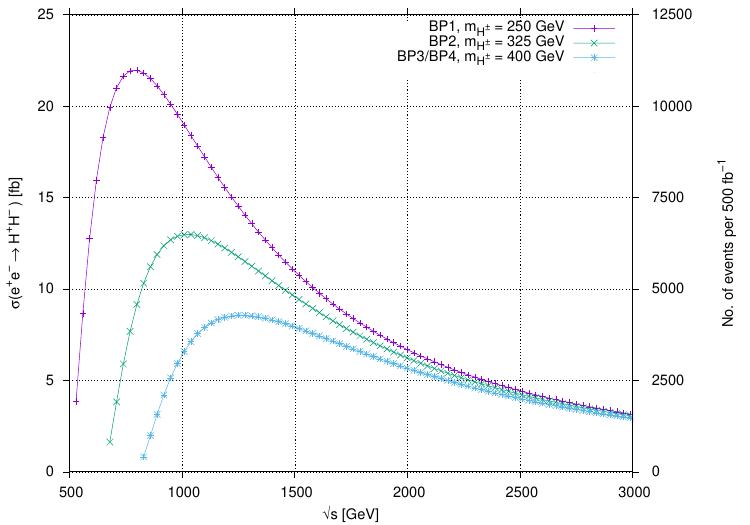}
\caption{\label{fig:1} The cross section versus energy at different benchmark points within the allowed region.}
\end{figure}

The  observability of the charged Higgs bosons is  investigated  through  signal  process  $ e^{+}e^{-}  \rightarrow   H^{+} H^{-}$.
The process chain is $ e^{+}e^{-} \rightarrow Z^{*}/\gamma^{*}\rightarrow H^{+} H^{-}\rightarrow H W^{+} H W^{-}\rightarrow   b \overline{b} jj b \overline{b} jj$. There are many other possibilities for this process. In the  process $ e^{+}e^{-} \rightarrow Zh $,   h  has taken as  SM higgs so it can not be taken for study when Higgs boson beyond SM is studied. In processes $ e^{+}e^{-}\rightarrow ZH $ or $ e^{+}e^{-} \rightarrow A h $, the  H Z Z  and    $ h Z A $  vertices are involved and both are  proportional to $ \cos\left( \beta-\alpha\right)  $. At  $\sin\left( \beta-\alpha\right)=1$, both processes vanish  when requirement is like SM~\cite{observe}. The choice of  process $e^{+}e^{-}\rightarrow Z^{*}/\gamma ^{*}\rightarrow H^{+} H^{-}$ is most suitable   and favorable for study of charged Higgs  at linear collider. The process takes place through exchange of Z and  gamma bosons in s-channel. Each heavy neutral Higgs boson $ H $ is allowed to  disintegrate only  into two b-jets and  W boson is allowed to decay into  two light jets.  The  final state contains  four b-jets, and fout light jets. \\
 In this study the masses of charged Higgs $ H^{\pm} $ and pseudo scalar Higgs A are considered same in order to avoid the decay of charged Higgs bosons into pseudo scalar Higgs A. The selected benchmark points which  fulfill the theoretical and experimental constriants are considered. The 2HDMC-1.7.0 \cite{hd} is linked with packages  HiggsBound-4.2.0 and HiggsSignal-1.4.0. It is used to make sure that all the benchmark points are consistent with all experimental and theoretical constraints. The range of parameter $m_{12}^{2}$ for each benchmark points satisfies the theoretical constraints. The mass splitting   between  heavy neutral Higgs and charged Higgs is taken in such a way that the bosonic decay of both charged Higgs $ H^{\pm}\rightarrow W^{\pm} H $ is kinematically permitted. The ranges of  heavy neutral and charged Higgs  bosons, $\tan \beta$, $m_{12}^{2}$ and $\sin (\beta-\alpha)$ for different benchmark points are listed in Table \ref{tab:2N}. 
\begin{table}[tbp]
\centering
\begin{tabular}{|c|c|c|c|c|c|c|}
\hline
&BP1&BP2&BP3&BP4 \\
\hline
 $ m_{h}$(GeV)& $125$& $125$ &$125$  & $125$  \\
\hline
 $m_{H}$(GeV)&$150 $&$200$ &$250$ &$300$  \\
\hline
$ m_{A}$ (GeV) & $ 250$ & $325$ & $400$ & $400$   \\ 
\hline
$ m_{H^{\pm}}$  (GeV)& $ 250$  & $325$ & $400$& $400$  \\
 \hline
$\tan \beta$&$40$&$40$&$40$&$40$ \\
\hline
  $ m_{12}^{2}$ & $560$& $996 - 999$ &$1560$  & $2245 - 2248$  \\
\hline
 $ \sin(\beta-\alpha)  $&$1 $&$1$ &$1$ &$1$  \\

\hline
\end{tabular}
\caption{\label{tab:2N} The Higgs bosons masses, range of $ m_{12}^{2} $ and branching fractions for different benchmark points within the allowed region.}
\end{table}
The branching ratios for decays $ (H^{\pm}\rightarrow W^{\pm}H )$ and $ (H\rightarrow b\overline{b}) $ as well as decay widths of higgs sector are obtained by using 2HDMC-1.7.0 \cite{hd} for selected benchmark points and are given in Table~\ref{tab:a1}. 
\begin{table}[tbp]
\centering
\begin{tabular}{|c|c|c|c|c|c|}
\hline
BPpoints&$ BR(H^{\pm}\rightarrow W^{\pm}H)$&$BR(H\rightarrow b\overline{b})$    \\
\hline
BP1 &9$.89\times10^{-1}$ &$7.112 \times10^{-1} $\\       
\hline
BP2&$9.95 \times10^{-1} $&$ 6.16\times10^{-1} $ \\
\hline
 BP3  &$ 9.97\times10^{-1} $&$ 5.08\times10^{-1} $   \\ 
\hline
 BP4 &$ 9.75\times10^{-1} $ &$ 3.79\times10^{-1} $ \\ 
\hline
\end{tabular}
\caption{\label{tab:a1} The branching ratios at  each benchmark points. }
\end{table}
All this information about a benchmark point is also included in its Susy LesHouches (SLHA) file and is provided to  Pythia8210 ~\cite{pyt} as input for generating events.
The cross section for the process at  given benchmark points is calculated  by using Comphep-4.5.2 ~\cite{cp,cf}   shown in Table~\ref{tab:3}. 
\begin{table}[tbp]
\centering
\begin{tabular}{|c|c|c|c|c|c|}
\hline
BP points&$\sigma$&$\sigma$& $\sigma$&$\sigma_{max}$&$\sqrt{s}$\\
&at$1$TeV&at$1.5$TeV&at$3$TeV&&\\
&(fb)&(fb)&(fb)&(fb)&GeV \\
\hline
BP1&17.294 &$10.978$&$3.1368$&$21.966$&$800$  \\       
\hline
BP2&11.699&$9.5886 $&$3.0445$&$12.977$&$1040$  \\
\hline
BP3&5.761&$7.9288$& $2.9294$&$8.5611$&$1250$   \\ 
\hline
BP4&5.761&$7.9288$&$2.9294$&$8.5611$&$1250$   \\ 
\hline
\end{tabular}
\caption{\label{tab:3} The cross section, maximum cross section and corresponding  C.M. energy of $e^{+}e^{-} \rightarrow  H^{+} H^{-} $ at different benchmark points. }
\end{table}
Energy is plotted versus cross section as shown in Figure ~\ref{fig:1}. 
 It shows that with increase in energy, the cross section increases rapidly. It has maximum value when energy  becomes equal to the threshhold energy. Threshold energy  is the energy, required to produce the process. The cross section decreases with further increase in energy and it almost becomes linear at 3 TeV. It is due to the reason that the cross section of a process is inversely proportional to the square of the center of mass energy. At 3 TeV center of mass energy, the crosssections for all benchmark points have the same value, it means  that the cross sections become independent of the center of mass energy and plots at this energy becomes linear."  

The  main SM background process in the signal channel is the production of pair of  top quark $t\overline{t} $ through electron-positron annihilation $ e^{+}e^{-} \rightarrow t \overline{t}$.
  Events for all this background processe are also generated in Pythia because all the known information about SM particles, their couplings and SM processes are already stored in Pythia built in flags.
\subsection{Event Selection process for identification  of jets}  
  
 The generated events are stored in HepMC-2.06.06~ \cite{ Hep}. The jets produced in the events are reconstructed by using the FASTJET ~\cite{ mm}. In this study anti-$k_{i}$, jet algorithm ~\cite{anti}  is used to reconstruct the jets. 
 Different kinematic selection cuts are applied which arises  certain fluctuations in the signal. The selection cuts are applied in such a way that enhances the signal to background ratio while signal events are maintained at an appropriate level. Selection efficiencies are calculated and  importance of signal is  determined by computing signal significance. These cuts define the band of ranges which are invariant quantities, measured in events. It must fulfil the number of several final state particles. These particles are identified in phase of primary reconstruction using “object identification cuts”. Then the kinematic selection cuts are applied to refine the rejection and selection of background events to finalise the results.\\

The first step in event selection is the kinematic cut on jets which omit the soft $p_T$ jets and the ones that are in forward region along the collision beams. For this we apply following cuts on transverse momentum and pseudorapidity of jets.
\begin{equation}
\label{eq:1}
p_{T}^{jet}  > 20 GeV , | \eta_{jets} | <  2.5
\end{equation} 
Once we have jets within the desired kinematic range, we split the reconstructed jets by identifying them as light and b-tagged jets. In order to achieve this, we do a $\Delta R$ matching of the jets with the generated particles which is defined as 
\begin{equation}
\label{eq:2}
\Delta R=\sqrt {(\Delta \eta)^{2}+(\Delta \phi)^{2}}
\end{equation}

We identify the jets which are within $\Delta R < 0.4$ of the b-quarks in the event as b-jets and the ones that are farther away from b-quarks as light jets. Once we have identified the jets, we apply the multiplicity cut on the jet. For $j\bar{j} j\bar{j} b \bar{b} b\bar{b}$ channel we require the event to have at least four light jets and at least four b-jets.
In the analysis, we use the selected light-jets (b-jets) to find a combination that minimise the $\chi^2$ defined as follows.

\begin{equation}
\chi^2 = \left(\frac{m_{jj,1} - m_W}{\sigma_{m_W}}\right)^2 + \left(\frac{m_{jj,2} - m_W}{\sigma_{m_W}}\right)^2 \left(\frac{m_{bb,1} - m_H}{\sigma_{m_H}}\right)^2 + \left(\frac{m_{bb,2} - m_H}{\sigma_{m_H}}\right)^2
\end{equation}  

where $m_{jj,bb}$ are the dijet mass, $m_W$ is the mass of $W$ boson, and $m_H$ is the mass of the heavy Higgs boson according to the BP taken, and the $\sigma_{m_{W,H}}$ are the widths of the respective mass distributions. The cut of $\chi^2 < 10$ is applied to select only events with good reconstructed $W$ and $H$ bosons. The charged Higgs boson $H^{\pm}$ is then reconstructed using the combination of $W$ and $H$ which gives mass nearest to $H^{\pm}$ nominal mass according to the BP. 
After applying different selections cuts,  The efficiencies are calculated.
For getting more better simulation results, more than hundred thousand events are generated and analyzed for each selected scenario. Applying all cuts on generated events,  relative efficiencies for  corresponding selection cut are calculated. At the end, total signal selection efficiency is also calculated for each benchmark point. The results are shown in Table~\ref{tab:9}.
\begin{table}[tbp]
\centering	
\begin{tabular}{|c|c|c|c|c|c|c|}
\hline
 Cuts & BP1 & BP2 & BP3 &BP4\\

 \hline
Four lightjets           &0.1467        &0.2557     &0.3278  &0.3052\\
\hline
   Four b-jets &0.5666       & 0.6478 &0.6655  &0.6544\\
     
\hline 
 $ \chi^{2} $&0.2224   &0.1737    &0.1085 & 0.6162   \\  
\hline
  CH $\chi^{2} $  &0.5027                      &0.4526     &0.3599  &  0.6767 \\ 
\hline 
 Total efficiency &    0.00929&0.01302&0.00852&0.00833  \\
 \hline
 $ \sigma \times B.R. $&4.58&2.17&0.66&0.37\\
 \hline
\end{tabular}
\caption{\label{tab:9}The efficiencies for different selection cuts at different mass hypothesis}
\end{table}
The signal process has reasonable selection efficiency at all the selected benchmark points. From Table~\ref{tab:9} it can be seen that from randomly generated signal events, all the events have four  b jets and light jets. This is due to the fact that  heavy scalar Higgs bosons, in signal processes  decay into b-jets and W bosons  decay into  light jets.
\section{Results and discussion}
\begin{figure}[h!] 
\centering 
\includegraphics[width=0.78\textwidth,clip]{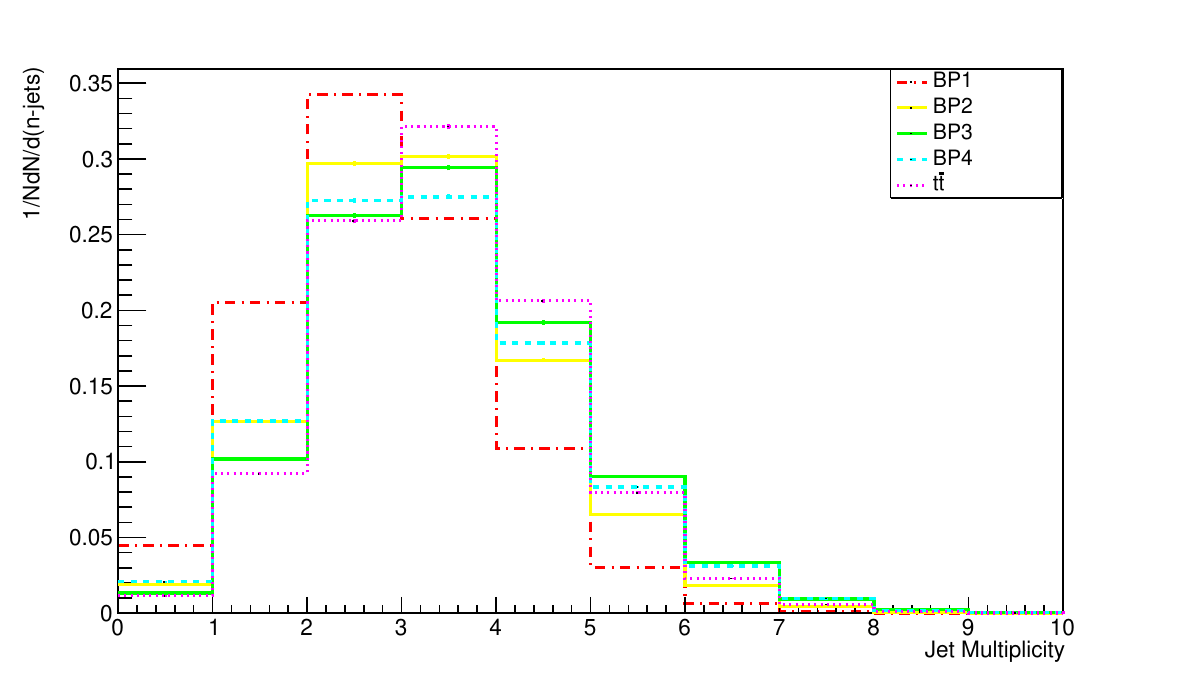}
\caption{\label{fig:2} The jet multiplicity for events(signal and background)}
\end{figure}

In the finalized results, topology of our assumed  first signal process contains four b-jets and four light jets.
 Jet multiplicity distributions of signals processes and different SM backgrounds are shown in Figure ~\ref{fig:2}.
The distribution of b-jets slightly depends on neutral Higgs boson mass in the assumed signal event. The production of b-jet is suppressed kinematically in the production of  Higgs boson, the availability of phase space is smaller due to which it decays to bottom quarks.
 
\begin{figure}[h!]
	
	\centering 
	\includegraphics[width=0.78\textwidth,clip]{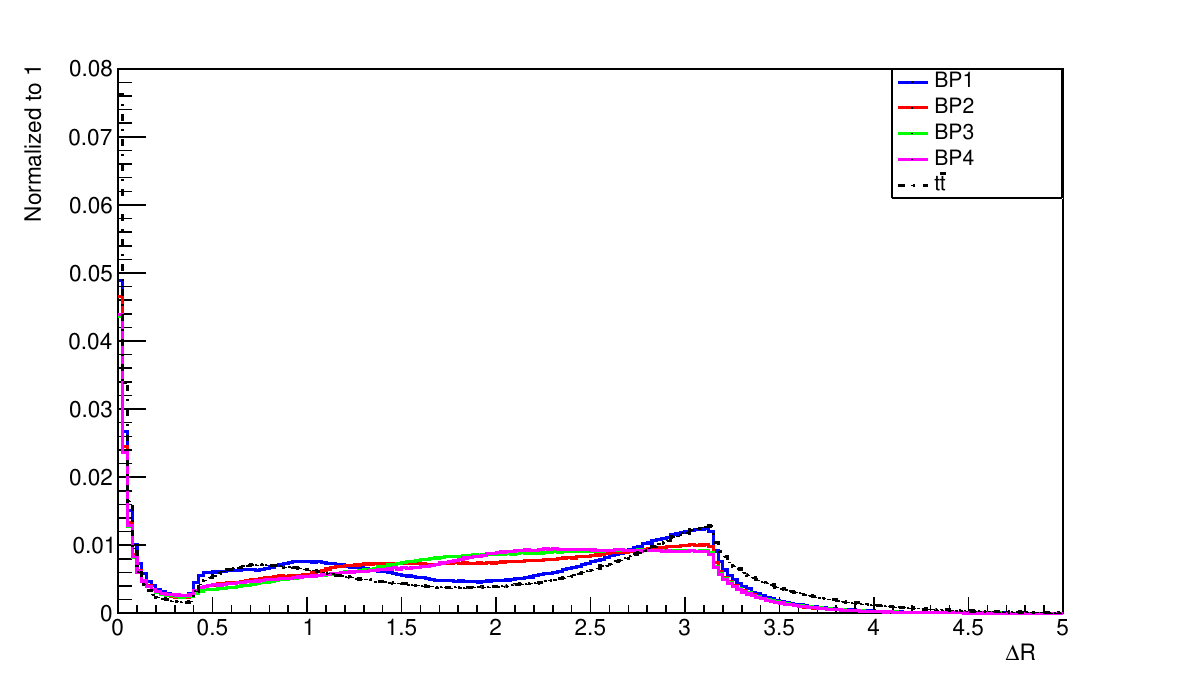}
	\caption{\label{fig:3} The distribution of  $ \Delta$ R( jets, quarks) variables used to tag bjet for all signal and background processes.}

\end{figure}

\begin{figure}[h!]
 	\centering 
 	\includegraphics[width=0.78\textwidth,clip]{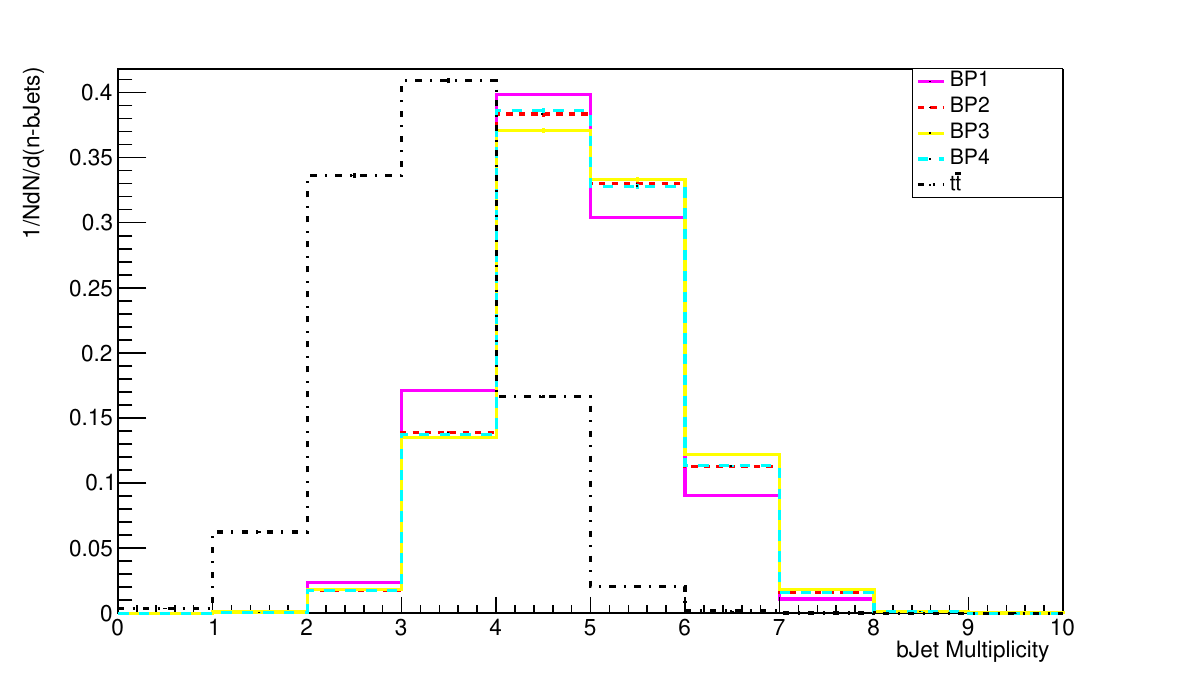}
 	\caption{\label{fig:4} The bjets multiplicity at different benchmark points in various signals and background events}
 \end{figure}

 Transverse energy  $E_{T}$ of jets for second process  is shown in Figure ~\ref{fig:1a}
 \begin{figure}[h!]
	\centering 
	\includegraphics[width=0.78\textwidth,clip]{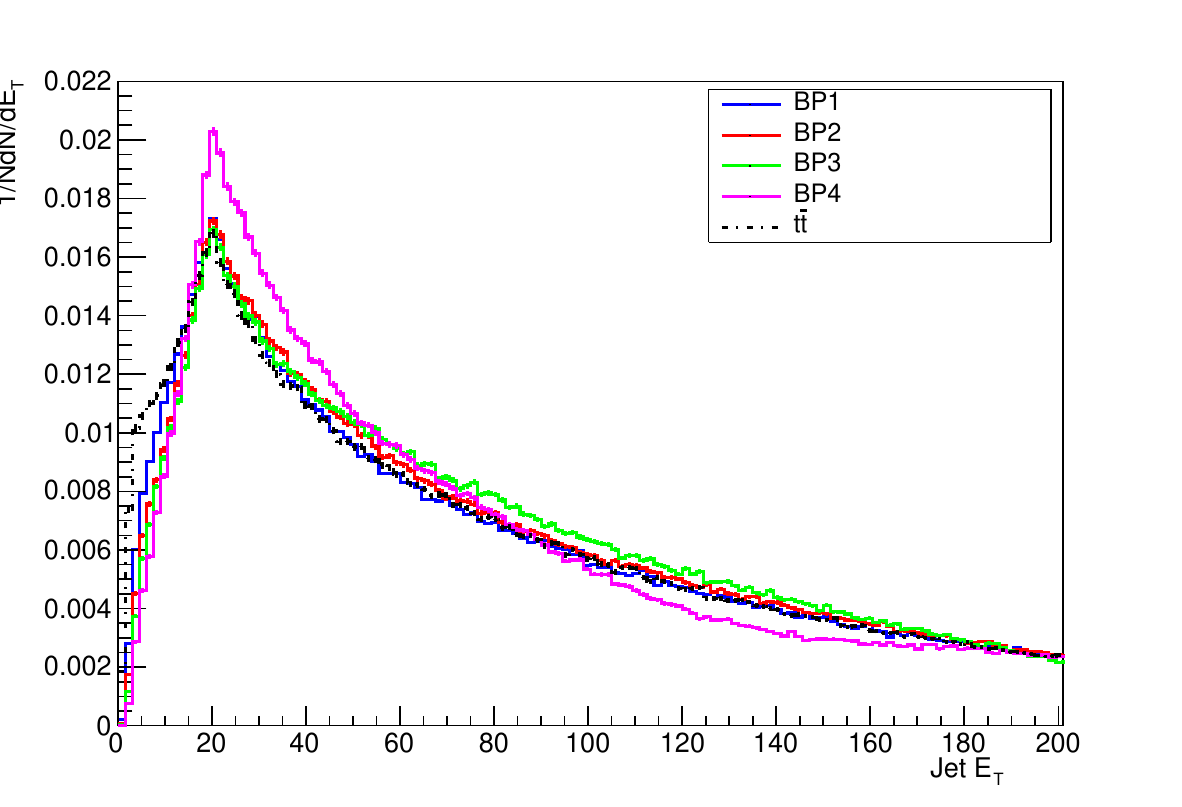}
	\caption{\label{fig:1a} The Pseudorapidity $\eta$ of signal along with all background events}
\end{figure}
The jet Pseudorapidity $ \eta $  for different signal and background processes is shown in Figure~\ref{fig:1a}.
\begin{figure}[h!]
	\centering 
	\includegraphics[width=0.78\textwidth,clip]{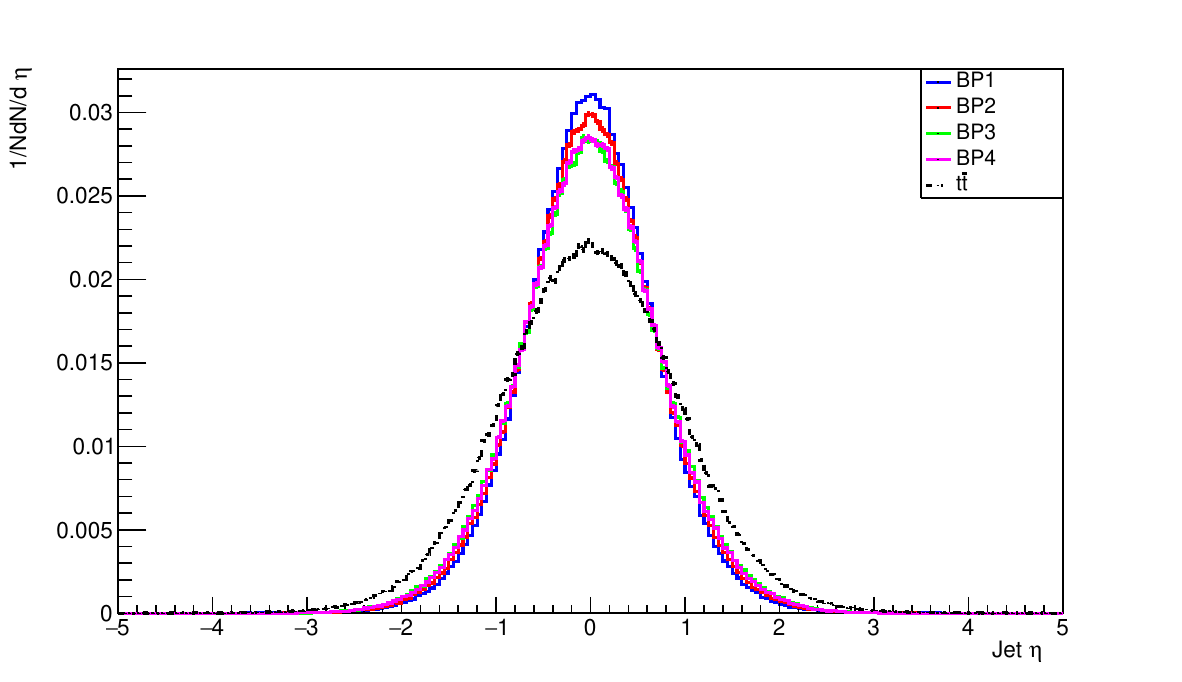}
	\caption{\label{fig:1a} The Pseudorapidity $\eta$ of signal along with all background events}
\end{figure}
After extracting out the data of $\Delta R$,  the profiling process of $\Delta R$ is discussed. By the analysis of plot of $\Delta R$ shown in Figure~\ref{fig:3},  the b-jets can easily be  identified by finding the minima of the plot. To identify the b-jets from all sorted jets, those jets are choosen which satisfy $\Delta R < 0.4$.  
  A jet is identified as a b-jet having possibility of $70\%$  if it has resemblance with a b quark  and with chance of $ 10\% $ if it resemblance with a c quark. The above mentioned values are supposed as the b tagging efficiency and fake rate successively. As signal comprises on four b-jets from the hadronic decay of each heavy Higgs boson denoted by $H_{1}$ and $H_{2} $. 
In the  Figure~\ref{fig:4} the number of b-jets in signals and background events are shown.

 After the selection cuts, the only background which can contribute up to a reasonable level is $ t\overline{t} $ having a very small number of events with four b-jets. For signal processes, it can be seen that each one has almost $60\% $to $70\% $ four b-jets efficiency. Other SM background may exist but they have very small number of events as compared to the signals.
 In the signal process, these jets come from hadronic decay of both W bosons. 

\subsection{Reconstruction  of W Bosons}
\begin{figure}[tbp]
	\centering 
	\includegraphics[width=0.78\textwidth.clip]{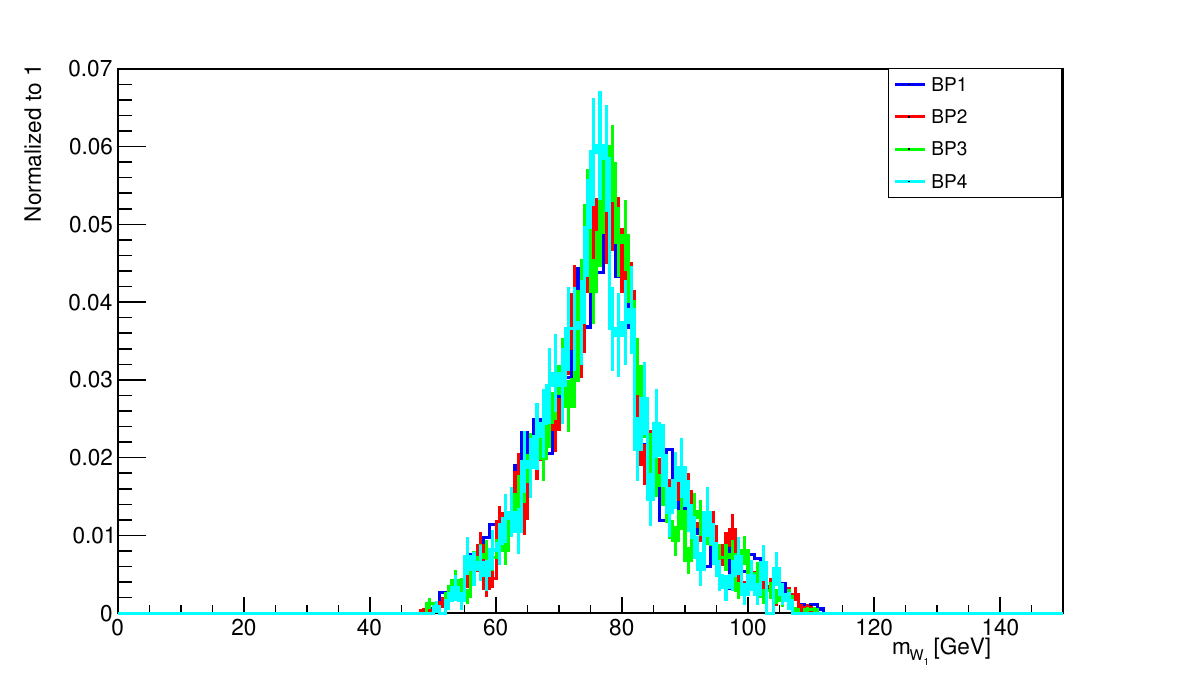}

	\caption{\label{fig:6} The plots of reconstructed invariant mass of$ W_{1} $ }
\end{figure}

\begin{figure}[tbp]
	\centering 
	\includegraphics[width=0.78\textwidth.clip]{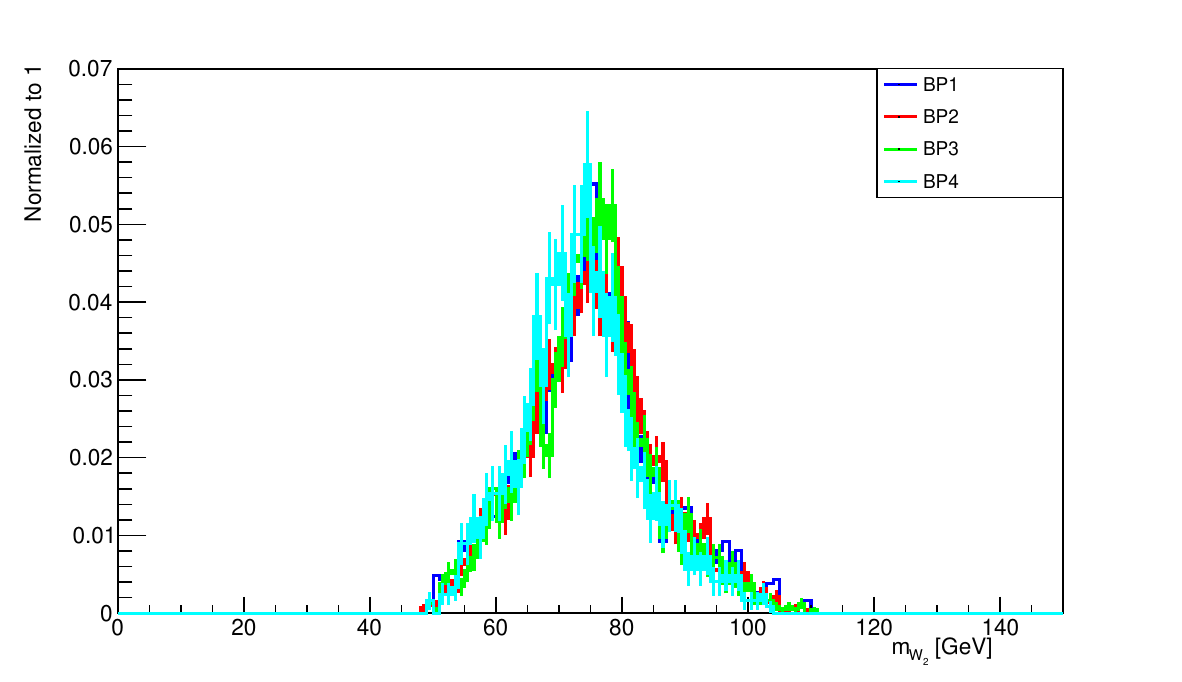}
	\caption{\label{fig:6a} The plots of reconstructed invariant mass of $ W_{2} $ in all signal events.}
\end{figure}

For all scenarios the reconstructed masses peaks are exactly at 80 GeV which is close to the known mass  of the W boson in SM (80.4 GeV). For backgrounds. the number of events with reconstructed W masses are very small as compared to our signals and on normalizing the above graphs, the reconstructed W masses for background processes cannot be seen.
The histograms are filled for reconstructed masses of both $ W^{\pm} $  for  all benchmark points and  shown in Figure~\ref{fig:6} and ~\ref{fig:6a}. The representation for $ W^{\pm} $ in histograms is taken as $ W_{1} $ and $ W_{2} $.
\subsection{Reconstruction  of Heavy Neutral scalar Higgs Bosons}
 The invariant mass of neutral scalar Higgs bosons is reconstructed. In this work, the generation of  signal and background processes is involved which display natural interference and selection techniques where different mass speculations are displayed for  Higgs invariant mass remaking.\\
The process of mass reconstruction of Higgs bosons is considered as  important to attain the reliable separation between the main assumed signal and the background processes. Peak of the signal resonance would be produced by the proper mass variable. By this process, large signal will be produced over the background ratio. The pair of b-jets comes out from the heavy Higgs Boson in the signal events. Due to that fact, the invariant mass of this pair should be lesser and lie within mass casement adjusted by neutral Higgs mass. Mass of the Higgs Boson can be  calculated by the conventional formula given as
\begin{equation}
\label{eq:9}
m_{H}=\sqrt {E^{2}-p_{x}^{2}-p_{y}^{2}-p_{z}^{2}}
\end{equation}

The reconstructed invariant mass of the Higgs boson is represented by "$m_{rec}$", and the actual value of the Higgs boson mass is represented by "$m_{Gen}$". Then, for each possible combination, the sum of squared differences between observed and predicted Higgs-boson masses is calculated. Then, light jet and bjet pairings that meet the following conditions are chosen.
\begin{center}

$\chi^{2}_{min}>10$
\end{center}
Only those events  are selected which have four b-jets. 
In $ \eta - \phi $ space, $\Delta R$ is calculated  for all probable combinations of b-jet pairs for each event. The selection cut is introduced to justify that the combination of b-jet pairs are truly coming from Higgs boson decay so their reconstructed masses should be nearly equal to input mass of heavy Higgs boson ($m_{H}$). The small fraction of reconstructed masses not lying in this window is eliminated from efficiency calculation for each signal.the representation of neutral Higgs bosons  are taken as $m_{H_{1}}$ and $m_{H_{2}}$.

\begin{figure}[tbp]
 \centering 
 \includegraphics[width=0.78\textwidth,clip]{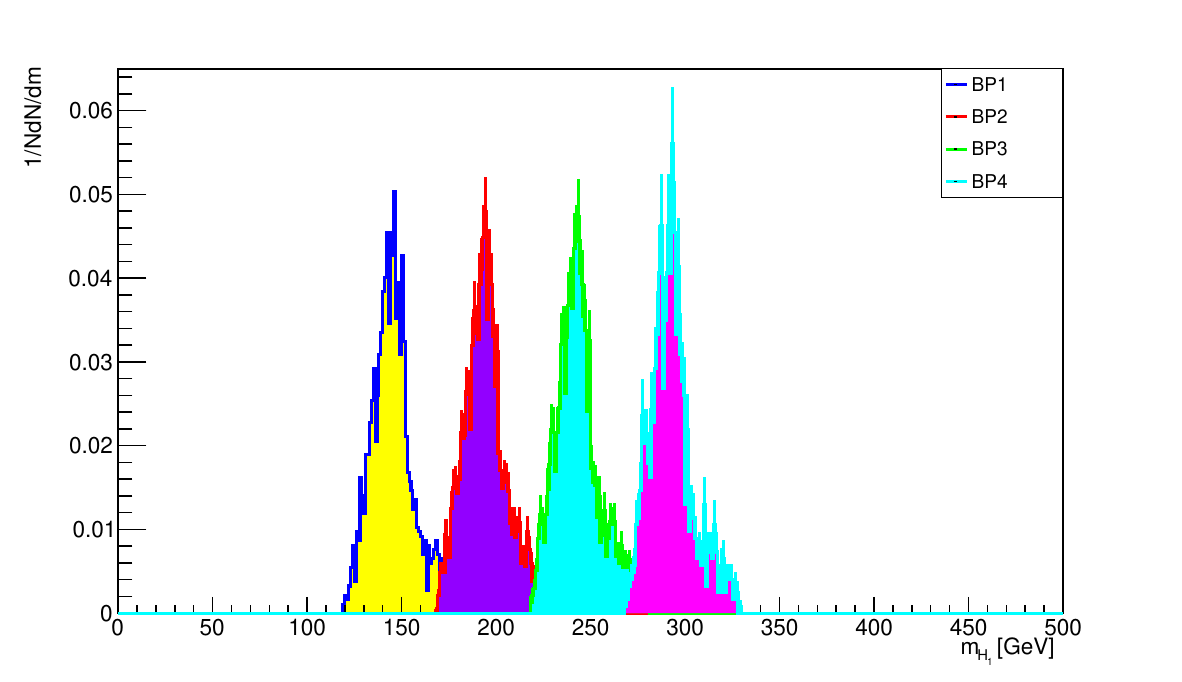}
\caption{\label{fig:8} The reconstructed invariant mass of first heavy Higgs boson mass $m_{H_{1}}$ for all signal events and  background.}
 \end{figure}

\begin{figure}[tbp]
 \centering 
 \includegraphics[width=0.78\textwidth,clip]{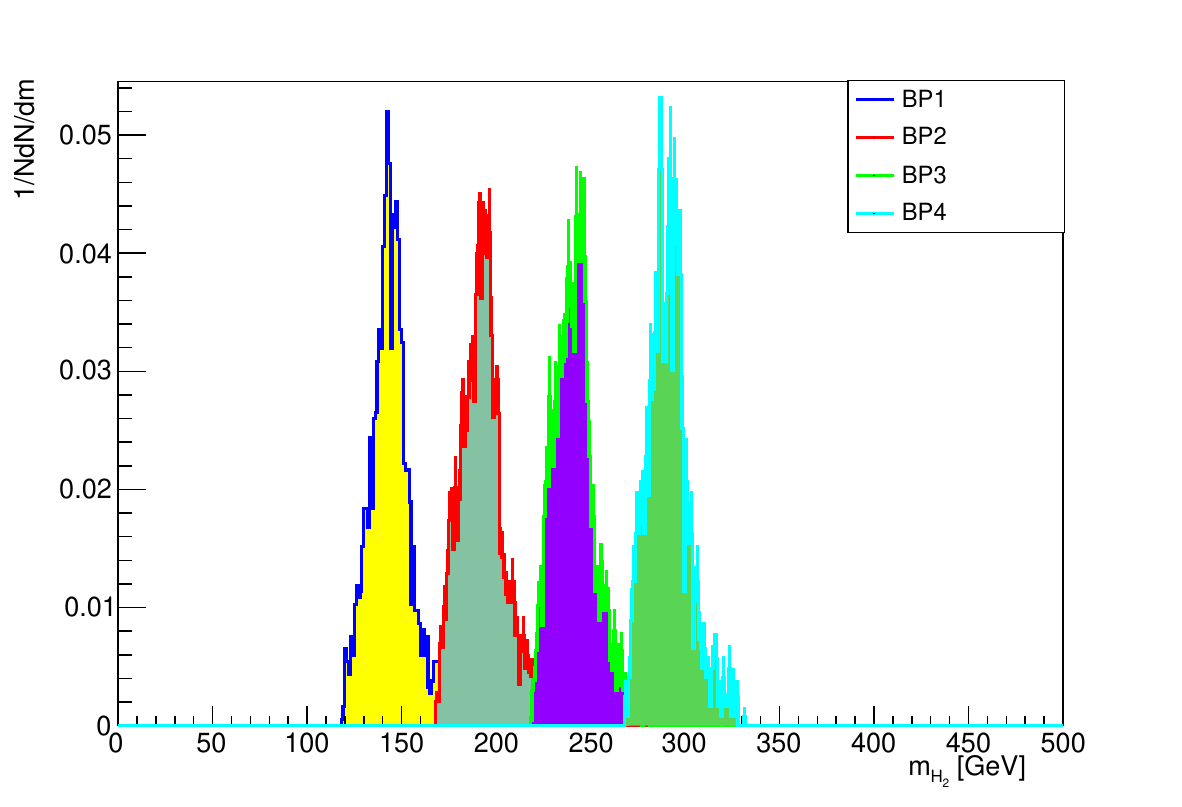}
\caption{\label{fig:9} The reconstructed invariant mass of first heavy Higgs boson mass $m_{H_{2}}$ for all signal events and  background.}
 \end{figure}

The reconstructed mass distributions of $H_{1}$ and $H_{2}$  are shown in Figures \ref{fig:8} and \ref{fig:9} accordingly.
The generated masses, reconstructed masses and corrected masses of Heavy neutral Higgs bosons for each selected scenario are computed in  Table~\ref{tab:5} and Table~\ref{tab:6}. 
\begin{table}[tbp]
\centering
\begin{tabular}{|c|c|c|c|}
 \hline	
 Signal Scenario & Gen. Mass(GeV)  & Recons. Mass (GeV) &Corr.recons. Mass(GeV)\\
  
\hline
BP1 & $150$& $143.9 \pm 0.2459$ &$151.45 \pm 0.49$     \\
\hline
BP2 & $200$& $193.42 \pm 0.2232$ &$200.97\pm 0.47$     \\
\hline
BP3 & $250$& $241.63 \pm 0.2476$ &$249.18 \pm 0.49$     \\
\hline
BP4 & $300$& $290.85 \pm 0.2920$ &$298.4 \pm 0.5$     \\
\hline
\end{tabular}
\caption{\label{tab:5}Generated mass, reconstructed mass and corrected reconstructed mass of heavy neutral Higgs boson $ H_{1} $ for all benchmark points. }
\end{table}
\begin{table}[tbp]
\centering
\begin{tabular}{|c|c|c|c|}
 \hline	
 Signal Scenario&Gen. Mass (GeV) & Recons. Mass (GeV) &Corr.recons. Mass(GeV)\\

\hline
BP1 & $150$& $143.24 \pm 0.24$ &$151.83 \pm 0.48$     \\
\hline
BP2 & $200$& $191.6 \pm 0.2245$ &$200.19\pm 0.47$     \\
\hline
BP3 & $250$& $240.8 \pm 0.2296$ &$249.39 \pm 0.48$     \\
\hline
BP4 & $300$& $290 \pm 0.2998$ &$298.59 \pm 0.5$     \\
\hline
\end{tabular}
\caption{\label{tab:6}The generated mass, reconstructed mass and corrected reconstructed mass of heavy neutral Higgs boson$ H_{2} $ for all benchmark points. }
\end{table}
The reconstructed masses are obtained by fitting suitable gaussian function on the distribution curves of $m_{H_{1}}  $ and $ m_{H_{2}} $ and taking their mean values. From Table~\ref{tab:5} and Table~\ref{tab:6} it can be observed that the reconstructed masses are less than the generated masses to some extent.
These errors are caused by the uncertainties in different stages of analysis process like jet reconstruction algorithm, b-jet tagging procedure, fit functions, momentum and energy calculations of particles etc. These uncertainties can be reduced by making further developments in jet cluster sequence, b-jet tagging algorithm, measuring and fitting methods etc. However, this study is not concerned with the implementation of such modifications. An average difference of all measured masses of $m_{H_{1}}  $ and $ m_{H_{2}} $ is determined from their generated values and it is found that these are 7.55 GeV and 8.59 GeV less from their generated values respectively and average mass error is 0.2521 and 0.2483 respectively. This average difference is added in the reconstructed masses of $m_{H_{1}}  $ and $ m_{H_{2}} $ to find the corrected reconstructed mass values. It can be observed that the corrected reconstructed masses of both heavy scalar Higgs bosons are well agreed with their generated masses.
\subsection{Reconstruction of  Charged Higgs Bosons}
The reconstruction of charged Higgs masses is possible from W and H bosons. Now those events are selected that have four b-jets as well as four light jets simultaneously. Mass of positively charged Higgs boson, expressed as $m_{H^{+}}$ is determined from its supposed decay products i.e. $W^{+}$ boson and one heavy Higgs boson  $m _{H_{1} }$. Mass of negatively charged Higgs boson expressed as $m_{H^{-}}$ is determined from its supposed decay products i.e. $W^{-}$ boson and other heavy Higgs boson  $m _{H_{2}} $. As the mass distributions of both heavy Higgs bosons are almost similar to each other so the choice of heavy Higgs boson to reconstruct the charged Higgs boson will not affect the charged Higgs rebuilt masses significantly.

\begin{figure}[tbp]
 \centering 
 \includegraphics[width=0.78\textwidth,clip]{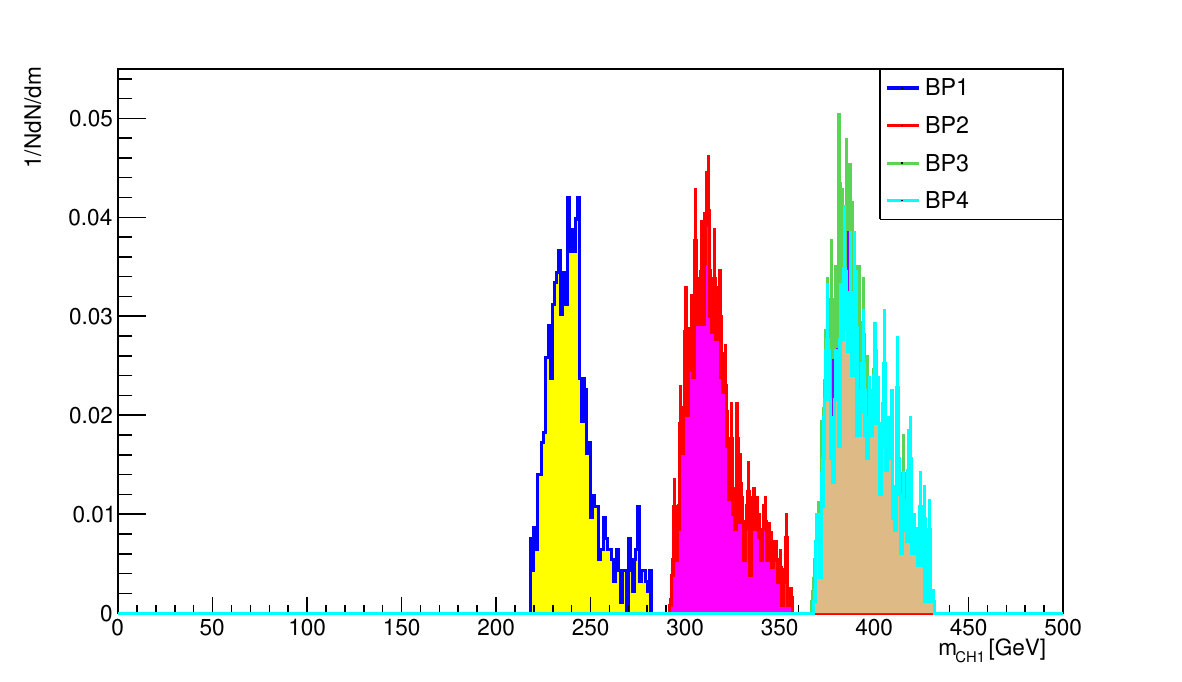}
\caption{\label{fig:10} The reconstructed mass of charged Higgs boson $ H^{+} $for all benchmark points.}
 \end{figure}

\begin{figure}[tbp]
 \centering 
 \includegraphics[width=0.78\textwidth,clip]{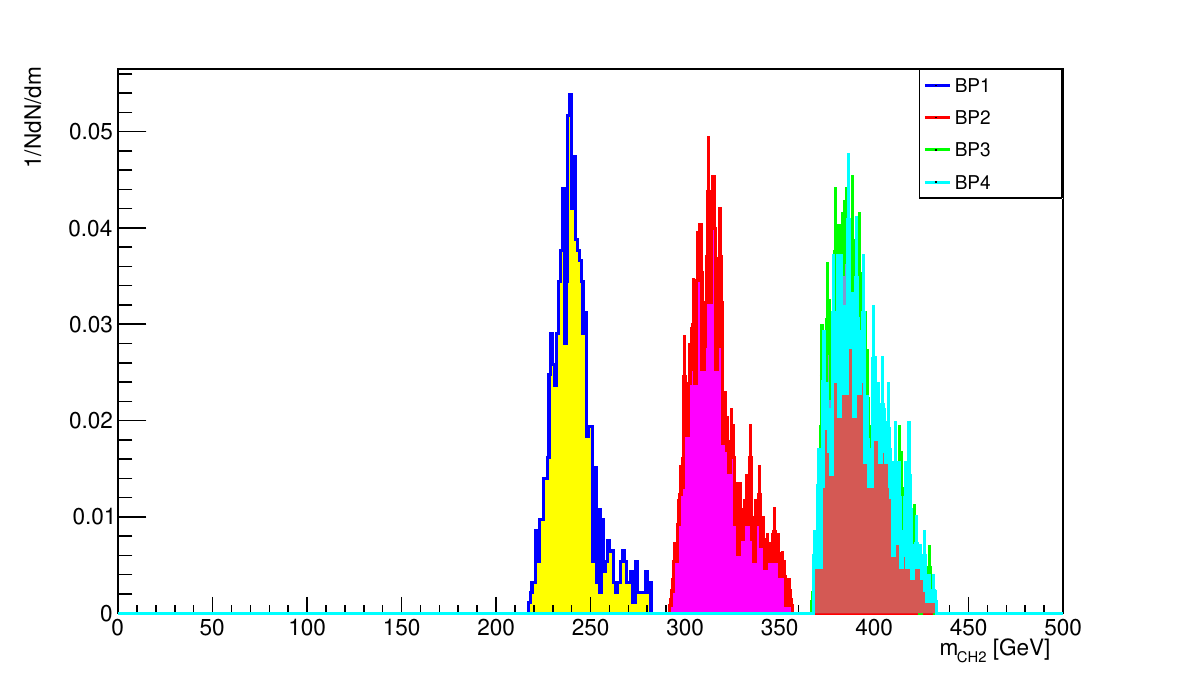}
\caption{\label{fig:10a} The reconstructed mass of charged Higgs boson $ H^{-} $for all benchmark points.}
 \end{figure}

 The  Figure~\ref{fig:10} shows the reproduced distributions for mass of charged Higgs  $ m_{H^{+}} $ and Figure~\ref{fig:10a}  for mass of charged Higgs  $ m_{H^{-}} $. The charged Higgs $ m_{H^{+}}$ is represented as CH1 and charged Higgs $ m_{H^{-}}$ is represented as CH2.
The masses of charged Higgs $ m_{H^{+}} $ and $ m_{H^{-}} $ are less from their generated mass by an average of 10.65 GeV and 11.31 GeV respectively and average mass error is 0.51 and 0.45 respectively.. These values are added in the reconstructed masses of $ m_{H^{+}} $ and $ m_{H^{-}} $ respectively to get corrected reconstructed masses. Table ~\ref{tab:7} and Table~\ref{tab:8} contain the data of generated, reconstructed and corrected reconstructed masses of  charged Higgs bosons. 
\begin{table}[tbp]
\centering
\begin{tabular}{|c|c|c|c|}
\hline	
Signal Scenario&Gen. Mass(GeV)&Recons. Mass (GeV) &Corr.recons. Mass(GeV)\\
 
\hline
BP1 & $250$& $237.5 \pm 0.4$ &$248.15 \pm 0.9$     \\
\hline
BP2 & $325$& $313\pm 0.4$ &$323.65\pm 0.9$     \\
\hline
BP3 & $400$& $388pm 0.6$ &$398.65 \pm 1$     \\
\hline
BP4 & $400$& $393.9 \pm 0.65$ &$404.55\pm 1$     \\
\hline
\end{tabular}
\caption{\label{tab:7}The generated mass, reconstructed mass and corrected reconstructed mass of charged Higgs boson $ H^{+} $ for all benchmark points.}
\end{table}
\begin{table}[tbp]
\centering
\begin{tabular}{|c|c|c|c|}
\hline	
Signal Scenario&Gen. Mass(GeV)&Recons. Mass(GeV)&Corr.recons. Mass(GeV)\\
\hline
BP1&$250$&$238.5\pm 0.29$&$249.81 \pm 0.7$     \\
\hline
BP2&$325$&$313.5\pm 0.4$&$324.81\pm 0.8$     \\
\hline
BP3&$400$&$386.8\pm 0.42$&$398.11 \pm 0.8$     \\
\hline
BP4&$400$& $390.95 \pm 0.7$&$402.26 \pm 1$     \\
\hline
\end{tabular}
\caption{\label{tab:8}The generated mass, reconstructed mass and corrected reconstructed mass of charged Higgs boson $ H^{-} $ for all benchmark points. }
\end{table}
 Gen. Mass is the mass of neutral heavy higgs mass taken as BP which satisfy the constraints. Here  it was generated .

\subsection{Signal Significance} 
\begin{figure}[h!]
 \centering 
 \includegraphics[width=0.78\textwidth,clip]{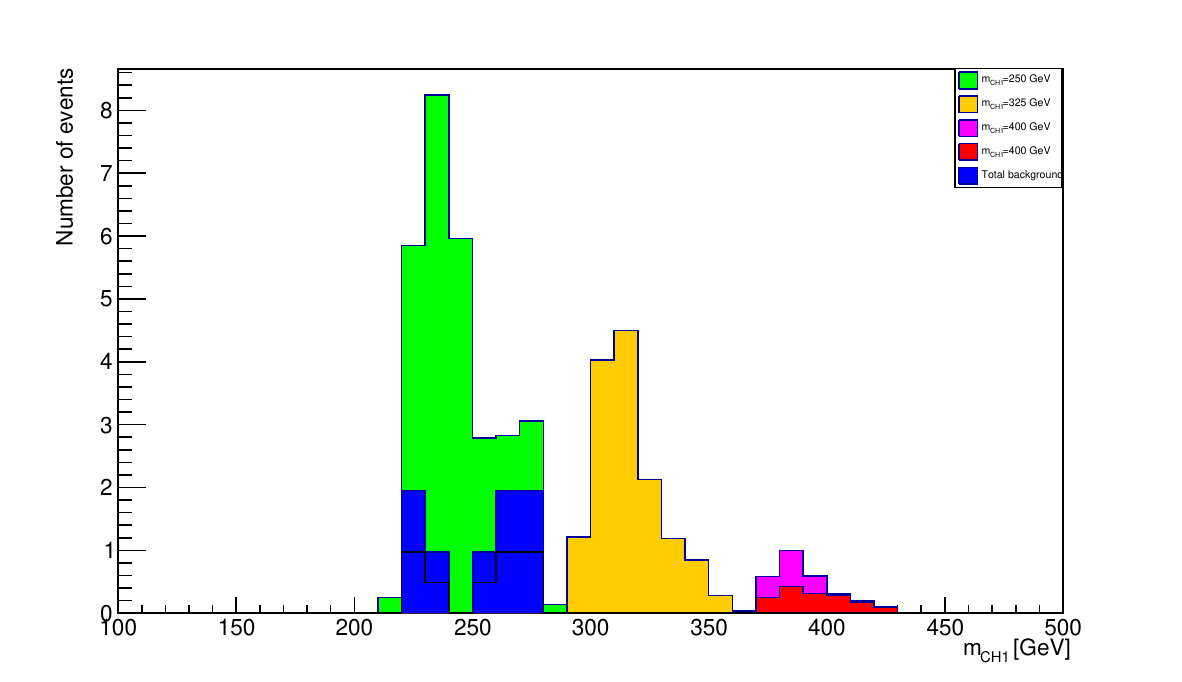}

\caption{\label{fig:13} Mass of charged Higgs boson for all signals and for total background events filled bin wise at $500 fb^{-1}$.}
 \end{figure}

\begin{figure}[h!]
 \centering 
 \includegraphics[width=0.78\textwidth,clip]{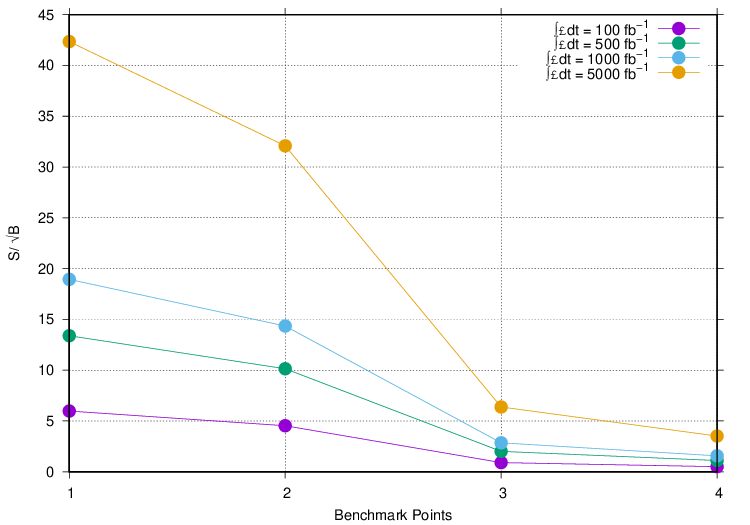}
\caption{\label{fig:14} Signal significance corresponding to each benchmark point at integrated luminosities of $100 fb^{-1}$, $500 fb^{-1}$, $1000 fb^{-1}$ and $5000 fb^{-1}$.}
 \end{figure}

\begin{figure}[h!]
 \centering 
 \includegraphics[width=0.78\textwidth,clip]{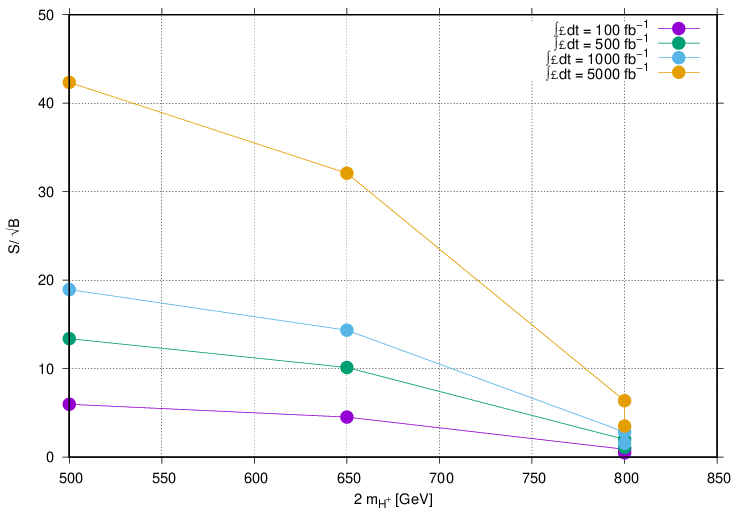}

\caption{\label{fig:15} The Signal significance versus twice of mass of  charged scalar  expected for $100 fb^{-1}$, $500 fb^{-1}$, $1000 fb^{-1}$ and $5000 fb^{-1}$.}
 \end{figure}

To examine the visibility of charged higgs boson at a linear collider,  significance of the signal is studied for charged Higgs mass distributions. Signal to background ratio, total efficiency  and signal significance is calculated for integrated luminosity 100  500 fb$^{-1}$ 1000 and 5000 and results is presented in Table ~ \ref{tab:10z} .
  Although in this study the detector effects are not included however this process can be used as a discovery channel for charged Higgs boson at CLIC. The Figure~\ref{fig:13} shows the bin wise filling of charged scalar Higgs mass values of all signal events and total background events at $100 fb^{-1}$, $500 fb^{-1}$.
 It can be seen that the signals are dominated over background events throughout the charged Higgs mass range.
 Table~\ref{tab:10z} shows the signal significance values at each benchmark point, at integrated  luminosities of $100 fb^{-1}$, $500 fb^{-1}$,$1000 fb^{-1}$ and $5000 fb^{-1}$. 
 \begin{table}[tbp]

 \centering	
\begin{tabular}{|c|c|c|c|c|c|c|c|}
 \hline
 & BP1 & BP2 & BP3 &BP4\\
 \hline 
 Significance $S/\sqrt{B}$  at$100 fb^{-1}$ & $5.98 $ & $ 4.53$ &  $0.90 $& $ 0.499 $  \\
 \hline
 
 Significance S/$\sqrt{B}$ at $500 fb^{-1}$& $13.39 $ & $ 10.14$ &  $2.01 $& $ 1.11$  \\
 
 \hline 
 Significance S/$\sqrt{B}$ at $1000fb^{-1} $& $18.94$ & $ 14.35$ &  $2.85 $& $ 1.57$  \\
 \hline
Significance S/$\sqrt{B}$ at $5000 fb^{-1}$ & $42.35 $ & $ 32.09$ &  $6.38$& $ 3.52$  \\
 \hline
Total Signal Efficiency $(\epsilon_{total})$& 0.00577&0.01302&0.00852&0.00833\\

\hline

\end{tabular}
\caption{\label{tab:10z} Values of signal significance and efficiency for all benchmark points at $100,500,1000$ and $5000 fb^{-1}$.}
\end{table}
  Figure~\ref{fig:14} shows the signal significance, against each benchmark point at integrated  luminosities, $100 fb^{-1}$, $500 fb^{-1}$,$1000 fb^{-1}$ and $5000 fb^{-1}$. Figure~\ref{fig:15} represents the signal significance  deviations from Standard Model predictions in the considered bosonic decay channel as a function of twice of the charged Higgs mass. The production of higher charged Higgs masses causes a reduction in the cross section which ultimately reduces the signal events at a specific value of integrated luminosity.

\section{Conclusion}
In this research work, 2HDM type-I is considered as theoretical ground and scenario chosen in it is like SM in which lighter scalar Higgs (h) behaves as Standard Model Higgs boson and $ \sin (\beta-\alpha) = 1$. Four distinct points in the allowed region are considered and their credibility is checked by 2HDMC-1.7.0. The production and observability of charged Higgs pair $ H^{\pm} $ through electron positron annihilation is investigated at four benchmark points  Compact Linear Collider. The series of decays in  signal process is given as $ e^{+}e^{-} \rightarrow Z ^{*}/\gamma^*\rightarrow H^{+} H^{-}\rightarrow H W^{+} H W^{-}\rightarrow jj b \overline{b} b \overline{b} jj$ in which bosonic decay of charged Higgs is considered that involves $H^{\pm} W^{\pm} H$ vertex twice in a signal process. This coupling vertex is proportional to $ \sin (\beta-\alpha) $ and momenta of charged Higgs and neutral scalar Higgs bosons. As value of $ \sin (\beta-\alpha)$ is set equal to unity so it is only proportional to the momentum of the particles involved.The value of $\tan \beta$ is kept relatively high to enhance the branching ratio $ H\rightarrow b \overline{b} $ to benefit the signal processes. \\
In this work a specific bosonic decay of charged Higgs boson is considered after its production which was not studied in detail till this time. The values of all Higgs masses are taken in such a way that they permit the assumed bosonic decay kinematically. The production cross section of the signal process is determined for each benchmark point at different center of mass energies which has reasonable values which shows that this process can be used to probe the charged Higgs boson experimentally. Ignoring the
minor errors, all the measurements for reconstructed charged and neutral Higgs boson invariant masses are in good agreement with their generated masses. Analysis code is run for all of assumed scenarios separately and the signal selection efficiencies are calculated for all benchmark points. The results shows that all the signals have sufficiently large total signal selection efficiencies. The reconstructed mass distributions of charged Higgs bosons and heavy Higgs bosons shows well high peaks. The analysis reveals that this process is favorable to discover the assumed scenarios for charged Higgs boson. Signal significance and total signal efficiency have been calculated for integrated luminosities at $100 fb^{-1}$ and $500 fb^{-1}$. The  results prove that the charged Higgs is observable through pair production along with its bosonic decays. This
study is supposed to provide the experimentalists a good way to examine the Higgs bosons beyond SM as well as to check the validity of 2HDM model in considered parameter space.


\end{document}